\documentclass[3p,times,procedia]{elsarticle}
\usepackage{nupha_ecrc}

\volume{00}

\firstpage{1}

\journalname{Nuclear Physics A}

\runauth{}

\jid{nupha}

\jnltitlelogo{Nuclear Physics A}

\usepackage{amssymb}

\usepackage[figuresright]{rotating}

\begin{document}

\begin{frontmatter}

\dochead{}

\title{Higher harmonic anisotropic flow of identified particles in Pb--Pb collisions with the ALICE detector}

\author{Naghmeh Mohammadi for the ALICE Collaboration}

\address{Nikhef, Science Park 105, 1098 XG Amsterdam, The Netherlands}

\begin{abstract}
Anisotropic flow plays a crucial role in establishing the equation of state of the Quark Gluon Plasma. The results at RHIC and LHC have demonstrated that the matter created in heavy-ion collisions behaves as a nearly perfect fluid reflected in the low value of the shear viscosity over entropy density ratio ($\eta/s$). The higher flow harmonics are particularly sensitive to the value of $\eta/s$ in hydrodynamic calculations. In this analysis, we present the first ALICE results on the $p_{\mathrm{T}}$ differential $v_{2}$, $v_{3}$ and $v_{4}$ for $\pi^{\pm}$, $\mathrm{K}^{\pm}$, p($\bar p$) from the high statistics 2011 heavy-ion run. We investigate how these $v_{n}$ coefficients evolve with particle mass and centrality for the 0--1\% and 20--30\% centrality ranges. These new measurements aim at differentiating between models that use different initial conditions, constraining further the value of $\eta/s$ and allowing to decouple the influence of the late hadronic stage from the hydrodynamic evolution of the system.
\end{abstract}


\end{frontmatter}

\section{Introduction}
\label{Introduction}
The main objective of heavy-ion collision experiments is to study the properties of the Quark Gluon Plasma \cite{Bass:1998vz}. Anisotropic flow, that characterises the momentum anisotropy of the final state particles, probes the properties of the system created in heavy-ion interactions, such as the equation of state and the ratio of shear viscosity to entropy density ($\eta/s$). This momentum anisotropy stems from the initial spatial anisotropy of the collision, reflected in its geometry, and by the initial density profile of nucleons participating in the collision which is different from one event to the other \cite{Alver:2008zza}. 
Anisotropic flow is characterized by a Fourier expansion of the azimuthal distribution of particle production relative to the symmetry plane \cite{Voloshin:1994mz}, according to:
\begin{equation}
dN/d(\varphi-\Psi_{n}) \approx 1+\sum_{n} 2v_{n}\cos[n(\varphi - \Psi_n)],
\label{particledistribution}
\end{equation}
where $\varphi$ is the azimuthal angle of particles, $\Psi_{n}$ the azimuthal angle of the nth-order symmetry plane and $v_{n}$ are the flow coefficients, with $v_{2}$, $v_{3}$ and $v_{4}$ 
 the elliptic, triangular and quadrangular flow, respectively. In this contribution, the results are reported for $v_{2}(p_{\mathrm{T}})$, $v_{3}(p_{\mathrm{T}})$ and $v_{4}(p_{\mathrm{T}})$ of identified charged pions, charged kaons and (anti-) protons in Pb--Pb collisions at $\sqrt{s_{\mathrm{NN}}} = 2.76$ TeV measured with the ALICE detector.
 \vspace{-0.25cm}
\section{Analysis details} 
\label{Analysis}

The data sample used for this analysis was recorded by ALICE during the 2011 Pb--Pb run at the LHC. Minimum-bias Pb--Pb events were triggered by the coincidence of signals from two forward detectors (V0A and V0C) positioned at both sides of the interaction point. In addition, an online selection based on the same V0 detectors was used to increase the number of central (i.e. 0--10$\%$ centrality range) and semi--central (i.e. 10--50$\%$ centrality range) events. \\

Events with a reconstructed primary vertex, whose position on the beam axis is within $\pm$10 cm from the nominal interaction point, were considered in this analysis. The data were grouped according to fractions of the hadronic cross section. The two centrality ranges presented in this article correspond to the 1\% most central (i.e.~smallest impact parameter) Pb--Pb collisions ($\sim$2.2 million) and the 20--30$\%$ interval (i.e.~semi-central with large impact parameter) with 1.5 million events. \\

The charged particle reconstruction was done with the Inner Tracking System (ITS) and the Time Projection Chamber (TPC).  The charged particle momenta were measured by the ITS and the TPC with a full azimuthal coverage in the pseudo-rapidity range $|\eta| < 0.9$. Charged pions and kaons as well as (anti-)protons were identified by combining the signals from the TPC and Time Of Flight (TOF) detectors similarly to the procedure followed in \cite{Abelev:2014pua}. In this analysis, pions and protons are identified with a purity of $>90\%$ in a transverse momentum range of $0.3<p_{\mathrm{T}}<6$ GeV/$c$ and kaons with a purity of $>85\%$ in $0.3<p_{\mathrm{T}}<4$ GeV/$c$. For more information about the ALICE detector and its performance see \cite{Abelev:2014ffa}.\\

The flow harmonics were measured with the scalar product method introduced in \cite{Sergei}. A pseudo-rapidity gap of $|\Delta\eta|> 0$ was applied between the identified pions, kaons and (anti-)protons selected from one side of the TPC and the charged reference particles selected from the other side of the TPC. Due to this small pseudorapidity gap applied, correlations not related to the common symmetry plane, known as non-flow, contribute to the measured $v_n^{\textrm{AA}}$ values. In this analysis, the contribution from non-flow was estimated using the HIJING event generator and is denoted as $\delta^{\mathrm{MC}}$ \cite{Sergei}. The maximum magnitude of this correction is around 0.017 for $v_2$, 0.008 for $v_3$ and 0.006 for $v_4$ in intermediate $p_{\mathrm{T}}$ region ($p_{\mathrm{T}}>3$ GeV/$c$). The reported values in this article are denoted as, $v_n^{AA-MC}$, and were extracted according to: 
\begin{equation}
v_{\textrm{n}}^{\textrm{AA-MC}} = v_{\textrm{n}}^{\textrm{AA}} -  \delta^{\textrm{MC}}
\label{estimated_corrected_vn}
 \end{equation}
More details about the applied non-flow subtraction can be found in \cite{Sergei}. 

\vspace{-0.35cm}
\section{Results} 
\label{Results}
\begin{figure}[!h]
\begin{center}
\includegraphics[width=0.325\textwidth]{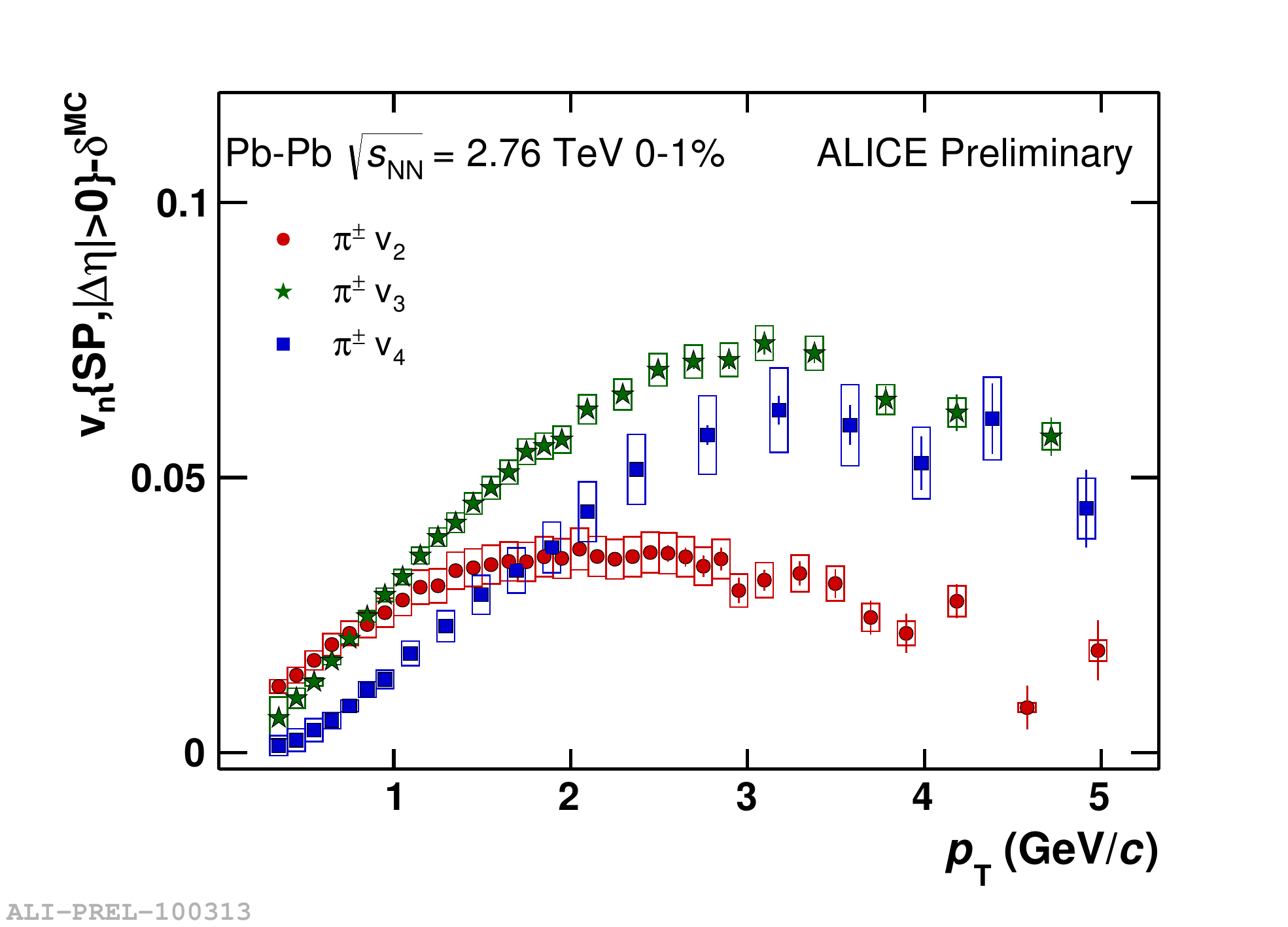}
\includegraphics[width=0.325\textwidth]{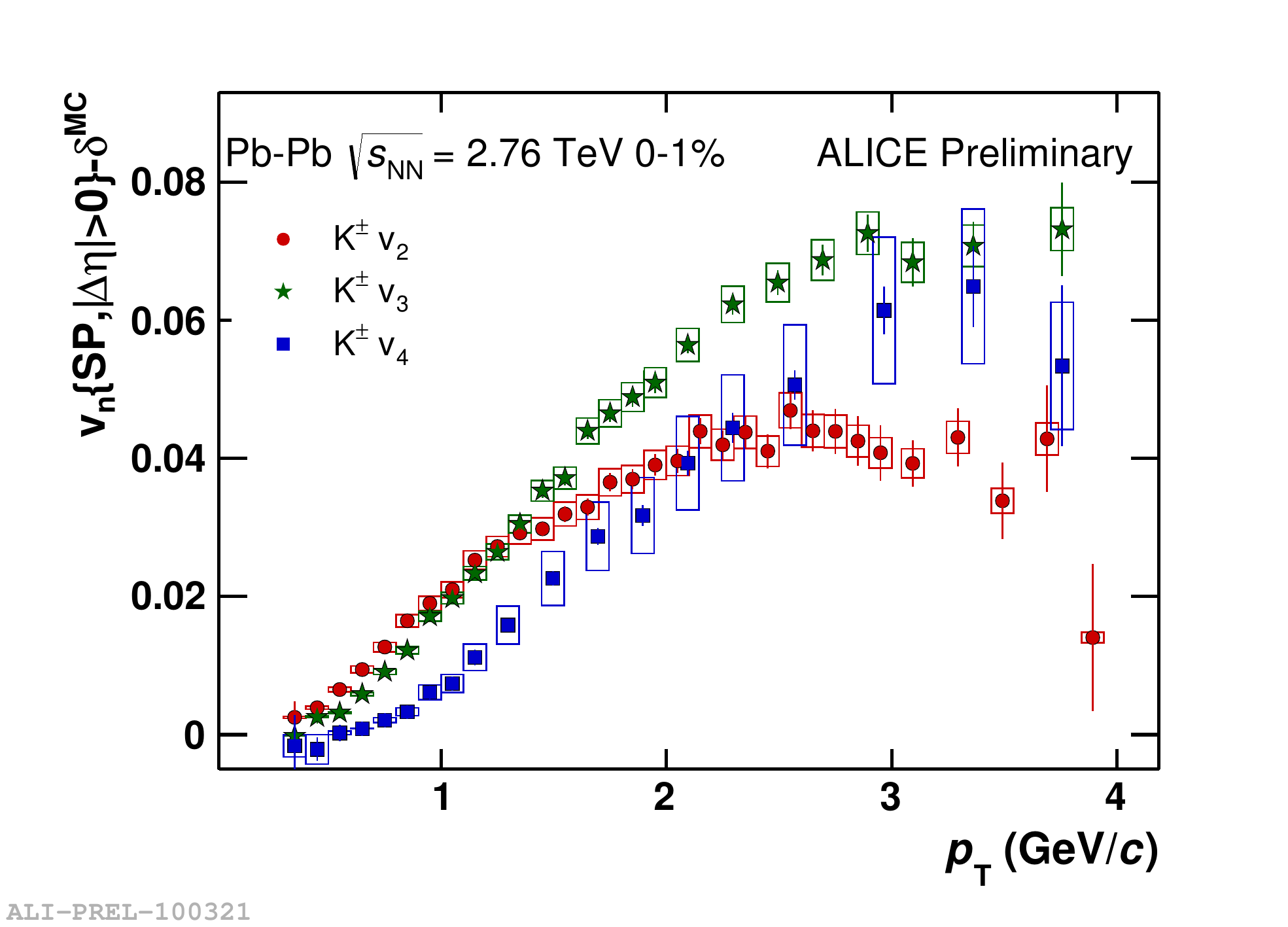}
\includegraphics[width=0.325\textwidth]{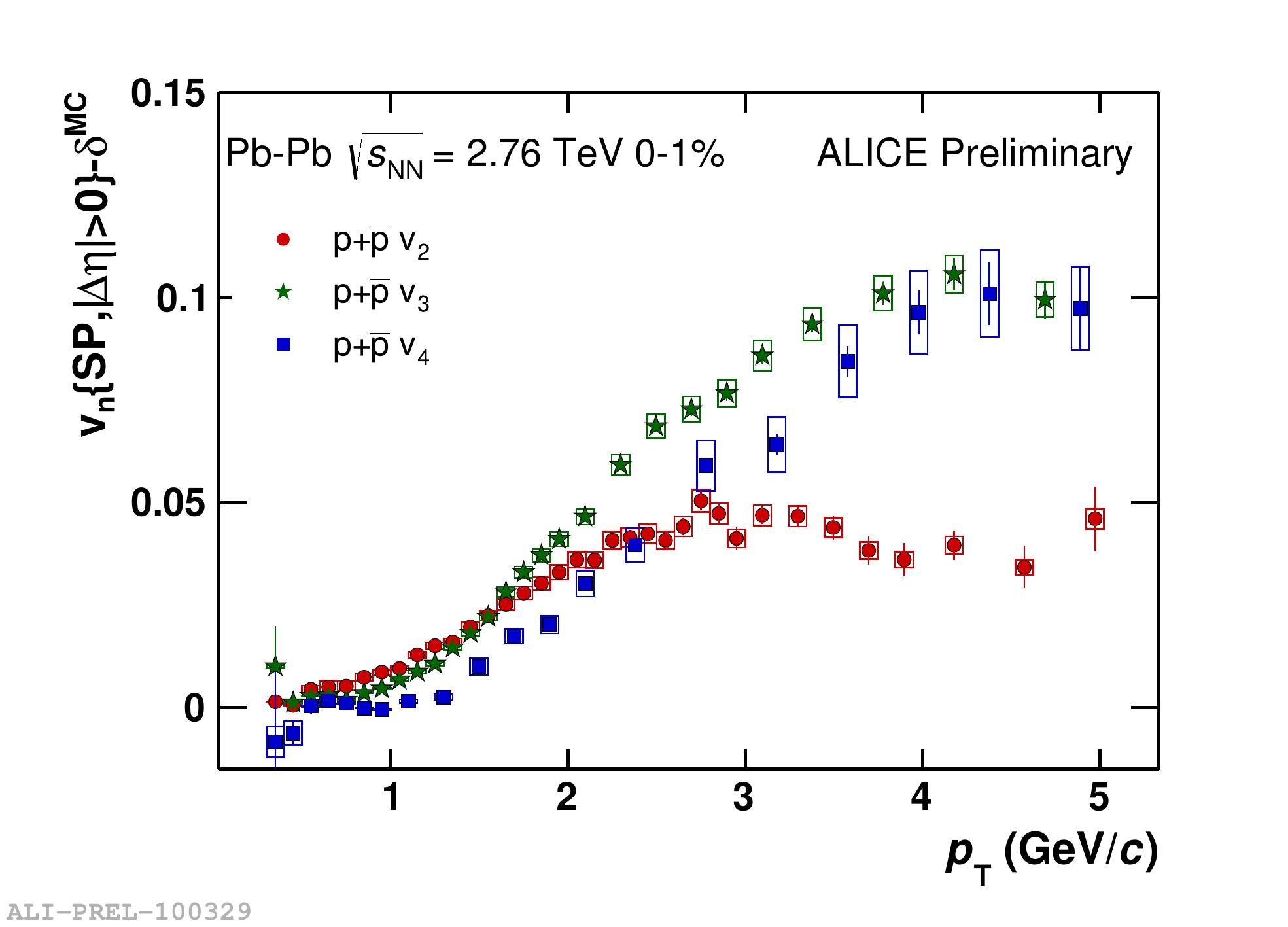}
\includegraphics[width=0.325\textwidth]{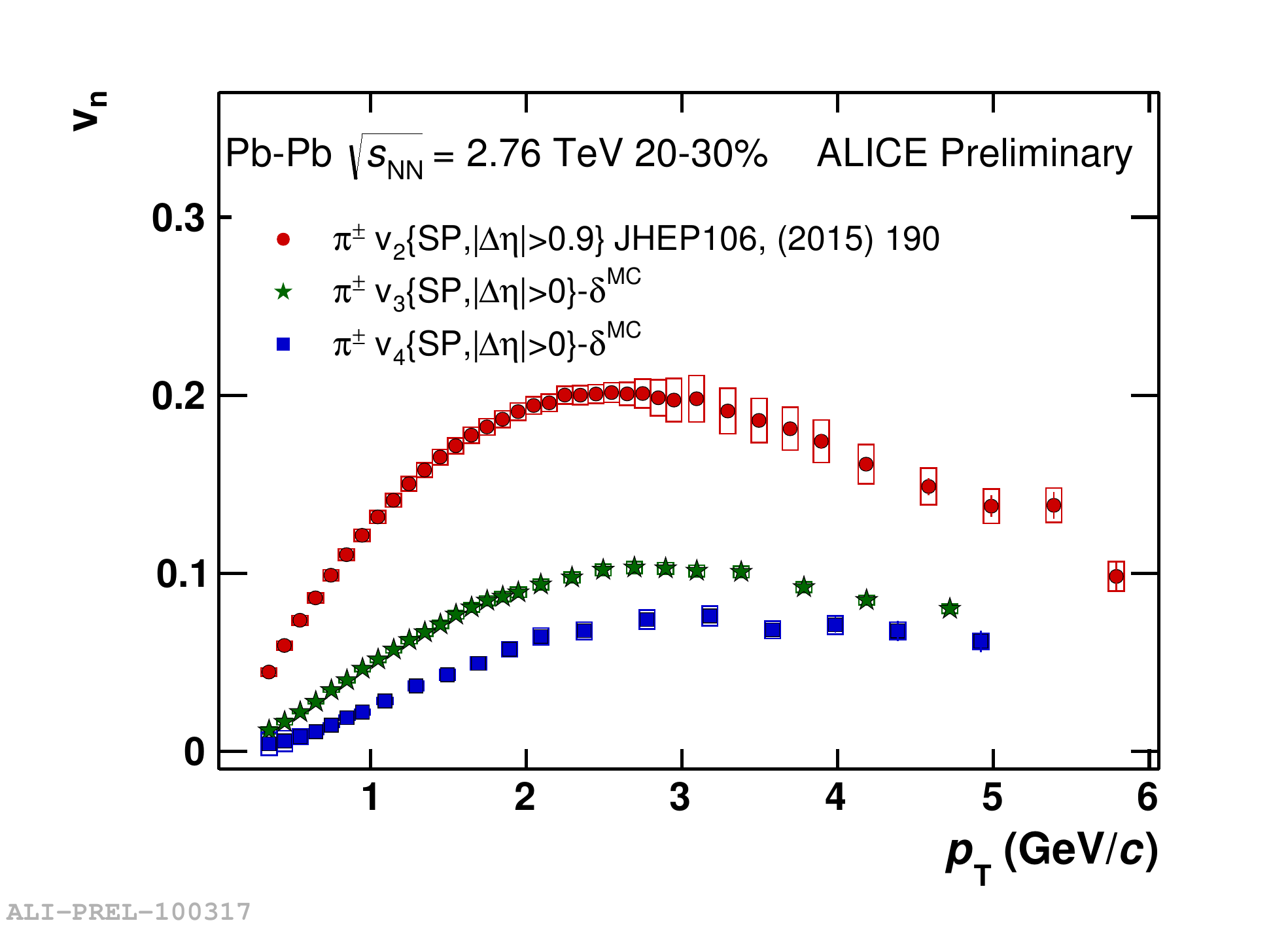}
\includegraphics[width=0.325\textwidth]{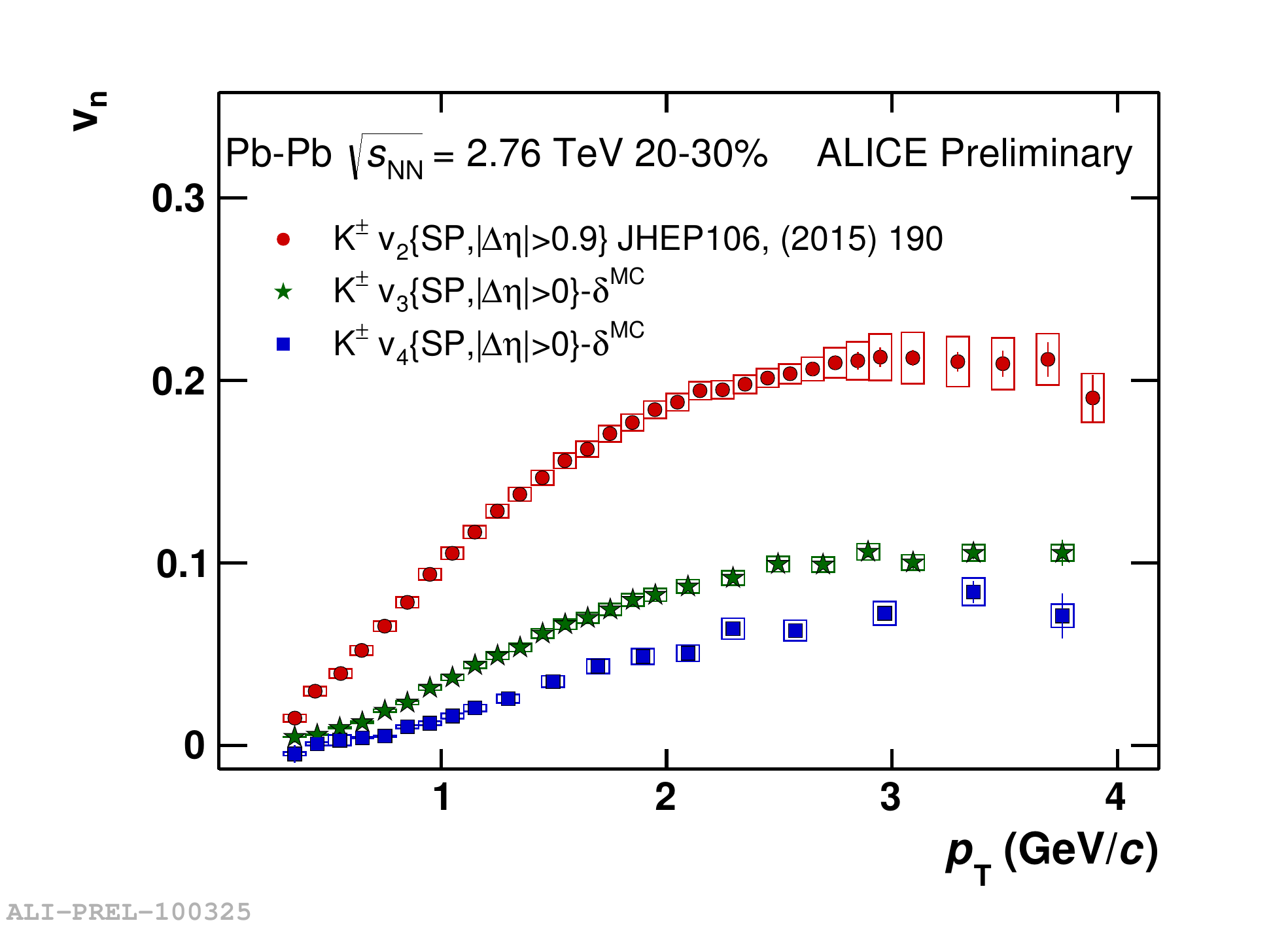}
 \includegraphics[width=0.325\textwidth]{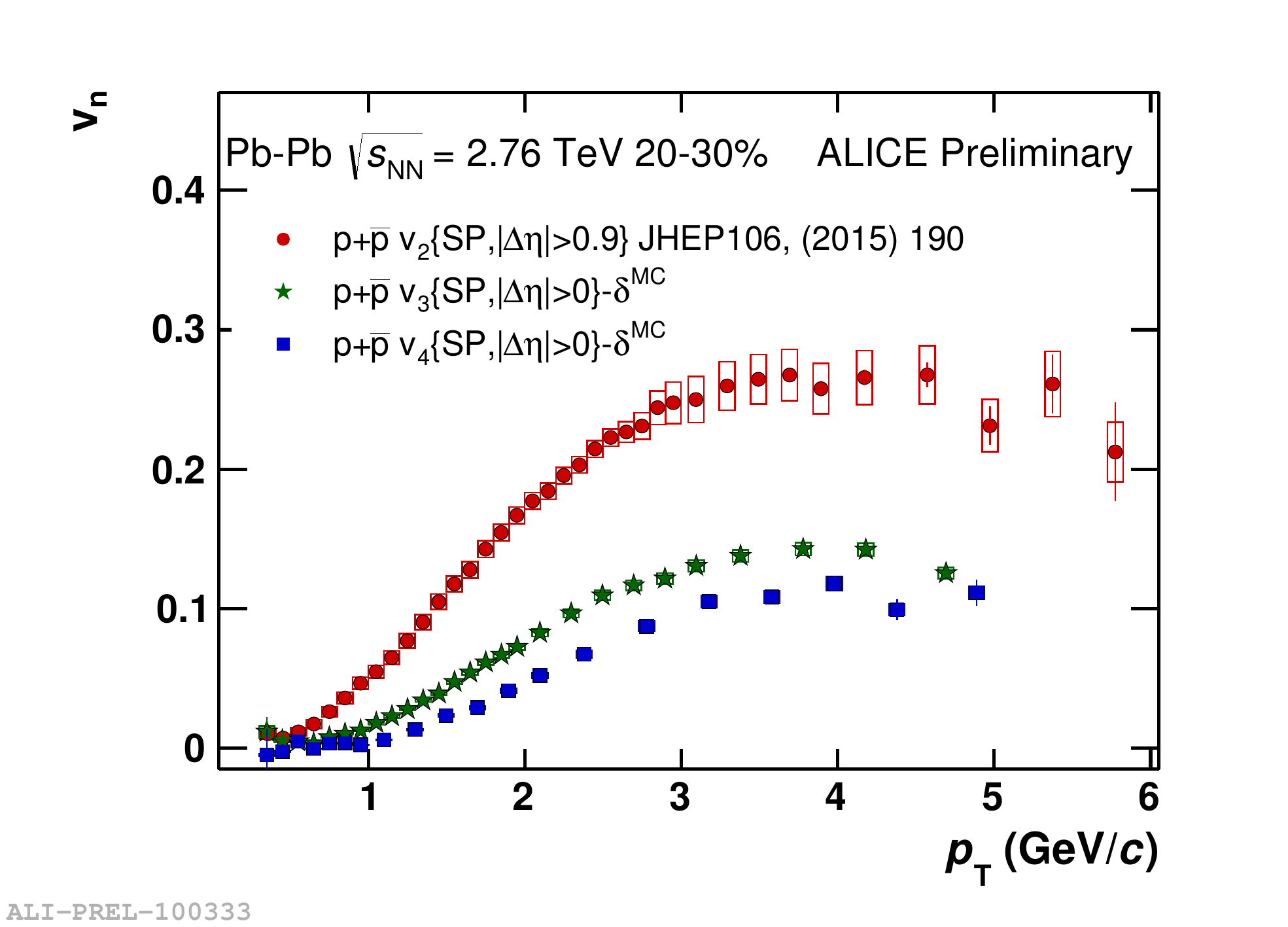}
\caption{The evolution of $v_2$, $v_3$ and $v_4$ with $p_{\mathrm{T}}$ for $\pi^{\pm}$ (left), $\mathrm{K}^{\pm}$ (middle) and $\mathrm{p}+\overline{\mathrm{p}}$ (right) for the 0--1$\%$ (top row) and 20--30$\%$ (bottom row) centrality ranges.}\label{vnEvolution}
\end{center} 
\end{figure}

Fig. \ref{vnEvolution} presents the $p_{\mathrm{T}}$ differential $v_{2}^{\mathrm{AA-MC}}$, $v_{3}^{\mathrm{AA-MC}}$ and $v_{4}^{\mathrm{AA-MC}}$ for charged pions and kaons and the combination of protons and anti-protons, in the left, middle and right plot, respectively, for the 1$\%$ most central Pb--Pb collisions at $\sqrt{s_{\mathrm{NN}}}=2.76$ TeV (upper row). The bottom row of Fig. 1 presents the same picture that develops for the 20--30$\%$ centrality range. Note that the $v_{2}$ values for this centrality range were calculated with the scalar product method with $|\Delta\eta|>0.9$ and without MC subtraction can be found in \cite{Abelev:2014pua}. \\ 

For the most central Pb--Pb collisions one expects the influence of the collision geometry to the development of $v_{n}$ to be reduced as compared to the contribution of the initial density fluctuations. It is seen that for charged pions around 0.8 GeV/$c$, $v_{3}$ gets equal to $v_{2}$ and becomes the dominant harmonic beyond this value. Furthermore, at around 2 GeV/$c$, $v_{4}$ becomes equal to $v_{2}$. For higher transverse momentum values, $v_{4}$ becomes gradually larger than $v_{2}$ reaching similar magnitude as $v_{3}$ at around 3.5 GeV/$c$. However, for the 20--30$\%$ centrality class where the collision geometry is expected to become the primary contributor to the development of $v_n$, it is seen that $v_2$ is the dominant harmonic throughout the entire measured momentum range. Furthermore, both $v_3$ and $v_4$ seem to have similar magnitudes and $p_{\mathrm{T}}$ evolution as the one observed for the most central Pb--Pb events, indicating a smaller influence of the collision geometry in their development than $v_2$.\\

For kaons and protons, one observes a similar trend in the $p_{\mathrm{T}}$-evolution of $v_2$, $v_3$ and $v_4$ as the one of charged pions. However, the flow harmonics exhibit a crossing that takes place at different $p_{\mathrm{T}}$ values than the ones for pions. For kaons, the $v_2$ and $v_3$ crossing happens at higher $p_{\mathrm{T}}$ ($\sim$1 GeV/$c$) compared to pions while for protons it happens at an even higher $p_{\mathrm{T}}$ value ($\sim$1.5 GeV/$c$). Similarly, the $v_2$ and $v_4$ crossing happens further in $p_{\mathrm{T}}$ for kaons ($\sim$2.4 GeV/$c$) and protons ($\sim$2.7 GeV/$c$) as compared to pions. The $v_4$ for kaons reaches similar magnitude as $v_{3}$ at around 3.5 GeV/$c$ and this takes place at around 4 GeV/$c$ for protons. The dependence of the crossing between different flow harmonics on the particle mass can be attributed to the interplay of not only elliptic but also triangular and quadrangular flow with radial flow. Consequently, the range where a given harmonic becomes dominant is shifted to higher $p_{\mathrm{T}}$ values with increasing particle mass.\\

\begin{figure}[!h]
\begin{center}
\includegraphics[width=0.325\textwidth]{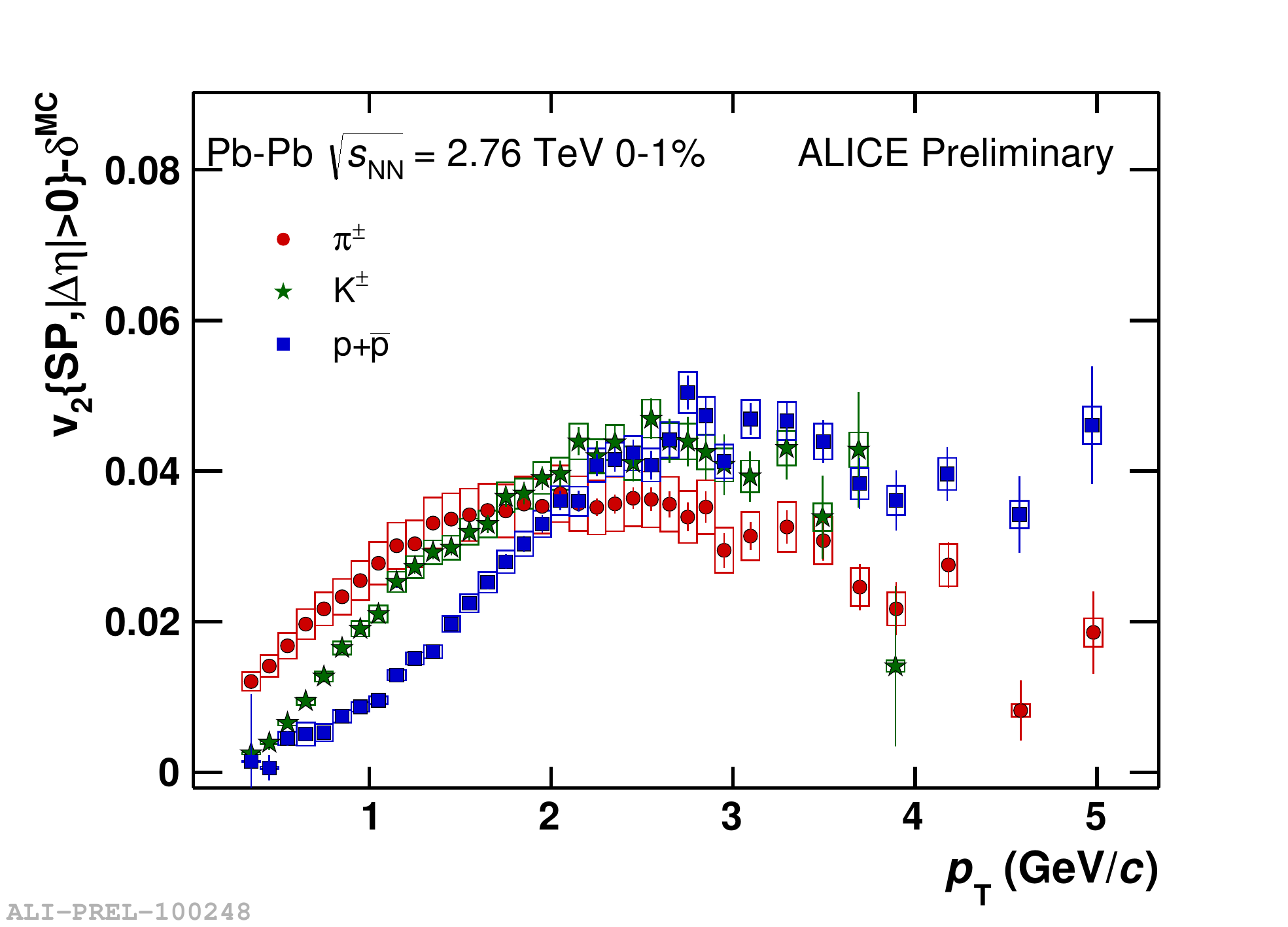}
\includegraphics[width=0.325\textwidth]{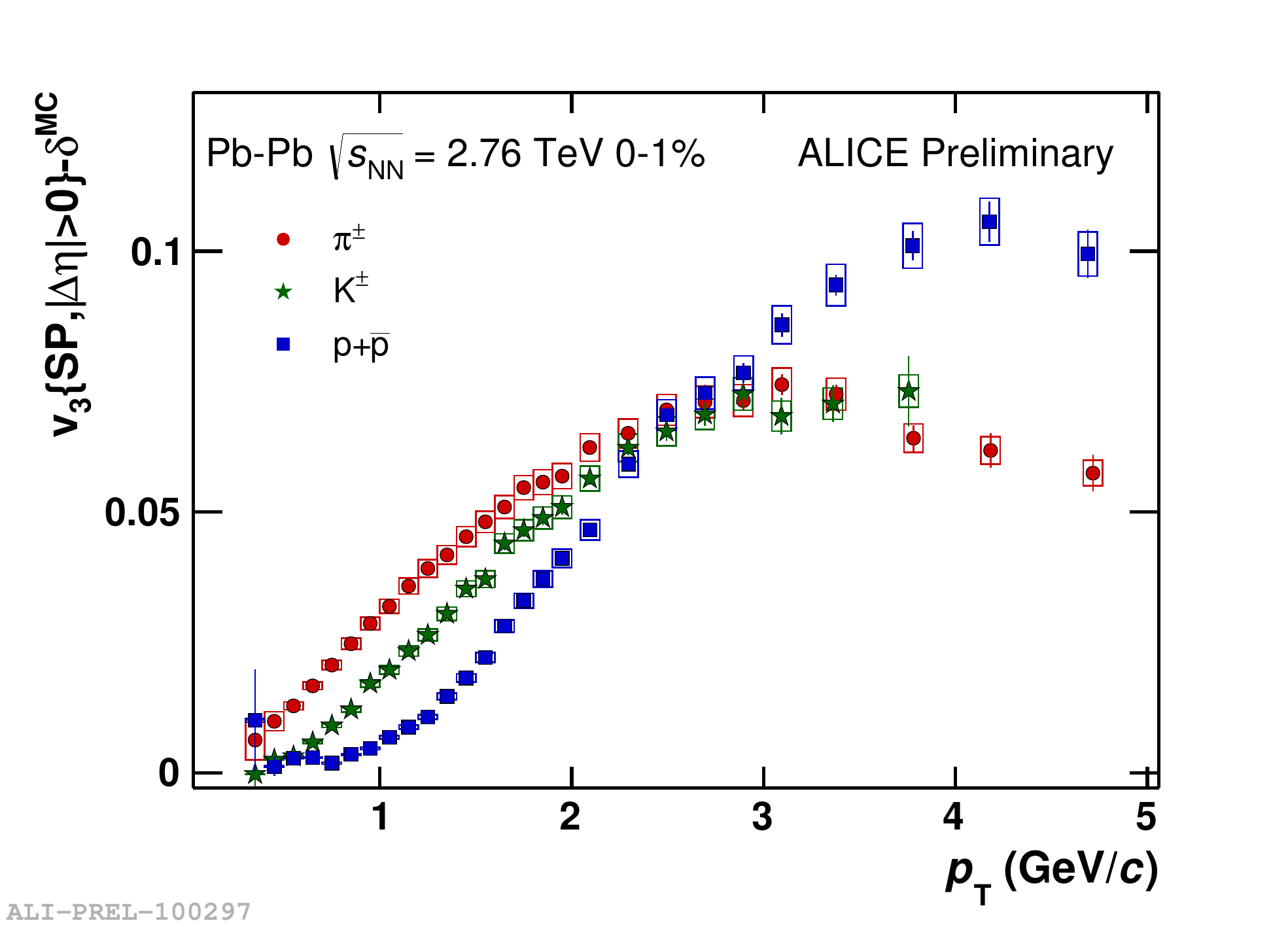}
\includegraphics[width=0.325\textwidth]{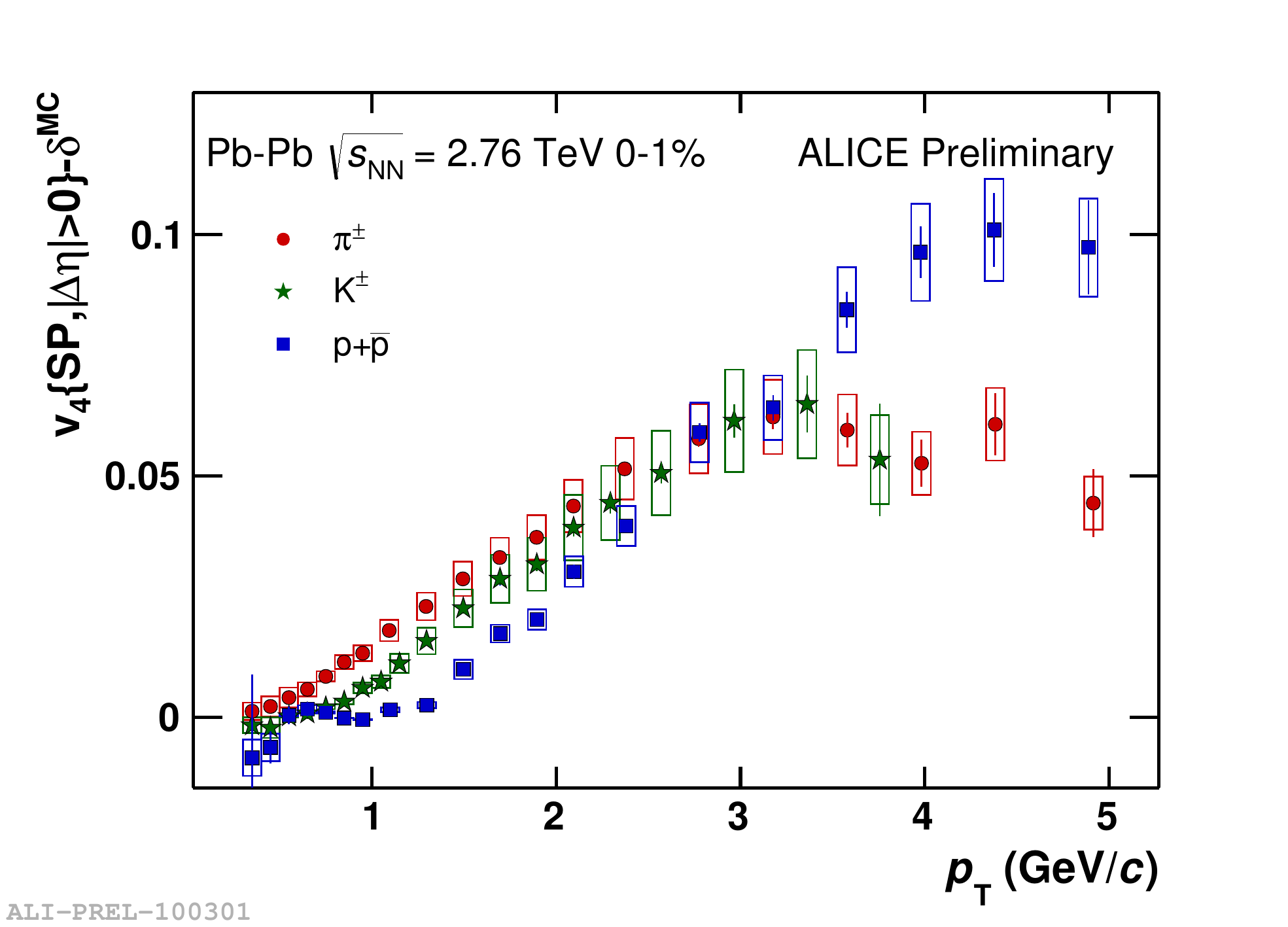}
\includegraphics[width=0.325\textwidth]{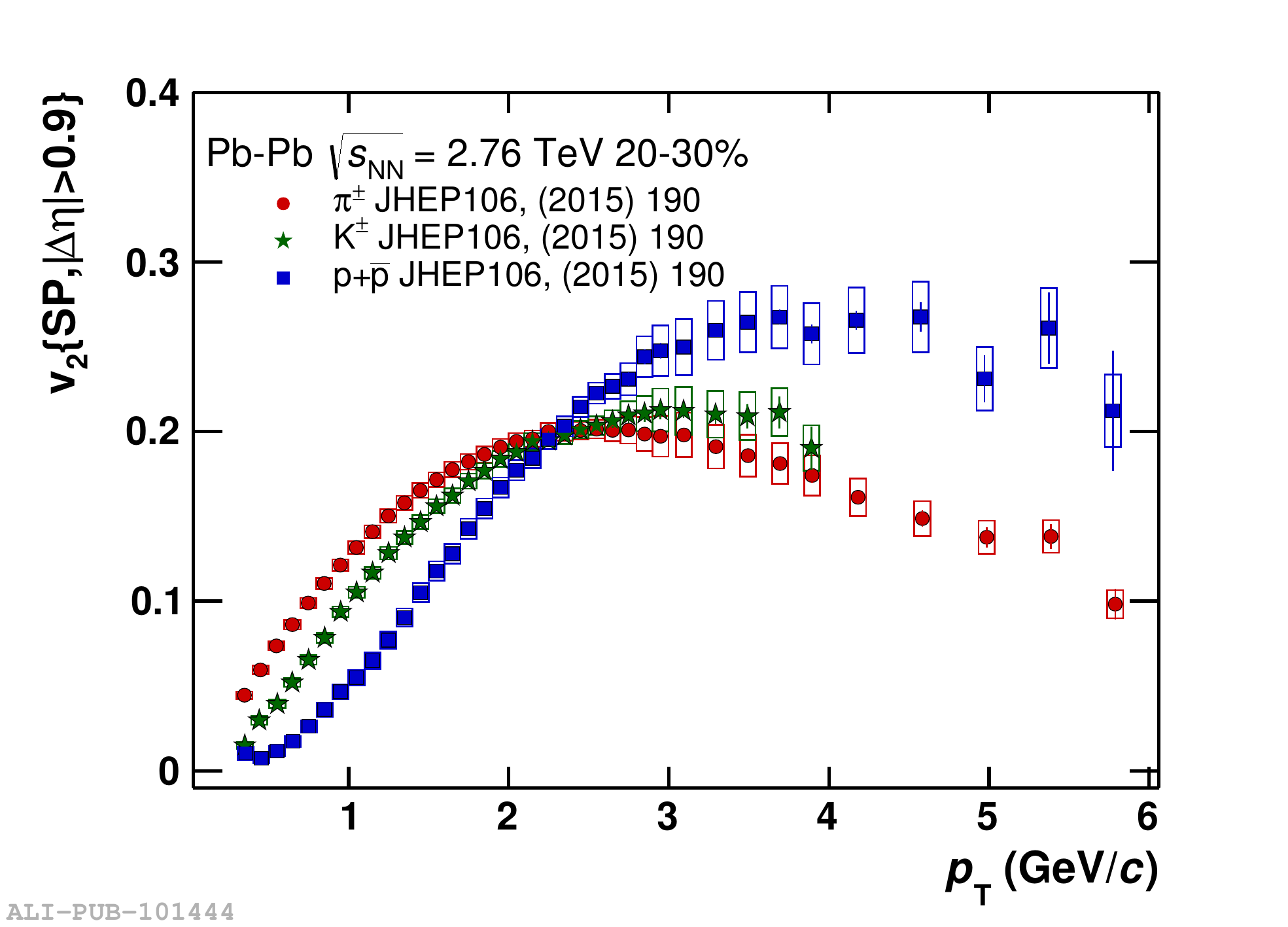}
\includegraphics[width=0.325\textwidth]{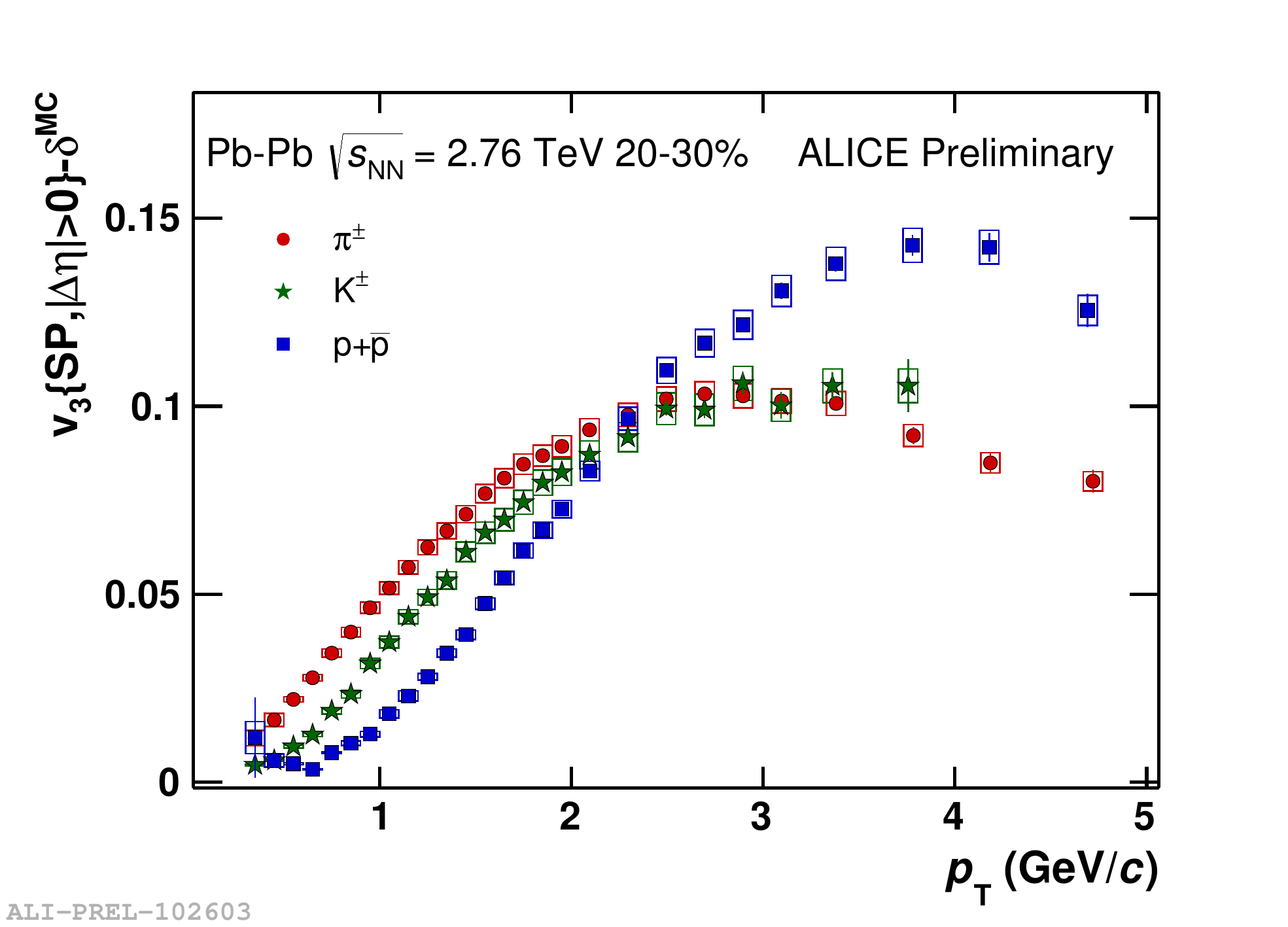}
\includegraphics[width=0.325\textwidth]{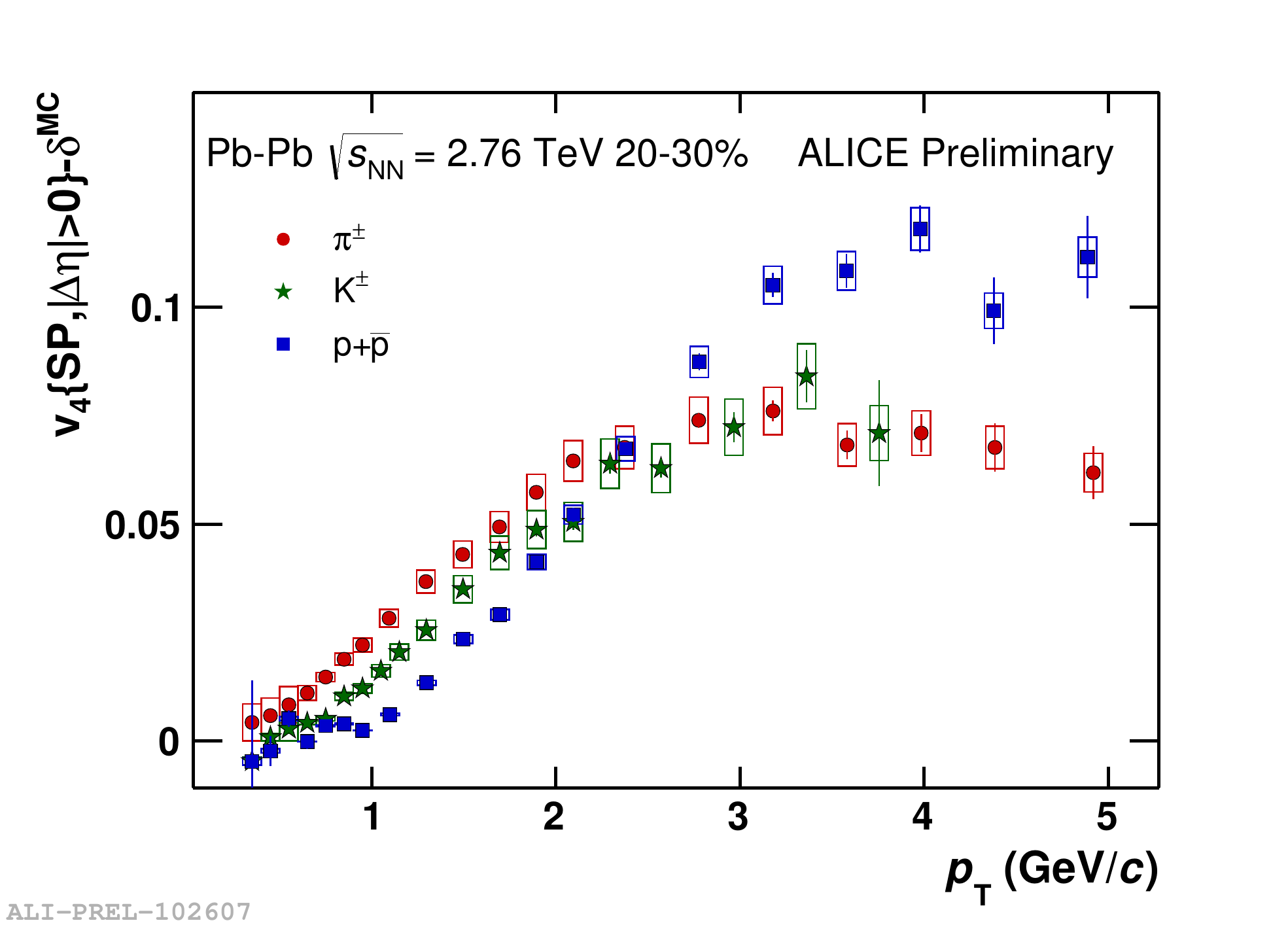}
\caption{The $p_{\mathrm{T}}$-differential $v_2$ (left) , $v_3$ (middle) and $v_4$ (right) for $\pi^{\pm}$, $\mathrm{K}^{\pm}$ and $\mathrm{p}+\overline{\mathrm{p}}$ for the 0--1$\%$ (top row) and 20--30$\%$ (bottom row) centrality ranges.}\label{massordering}
\end{center}
\end{figure}

Figure \ref{massordering} presents the $p_{\mathrm{T}}$ differential $v_{2}^{\mathrm{AA-MC}}$ (left), $v_{3}^{\mathrm{AA-MC}}$ (middle) and $v_{4}^{\mathrm{AA-MC}}$ (right) for $\pi^{\pm}$, $\mathrm{K}^{\pm}$ and $\mathrm{p}+\overline{\mathrm{p}}$ at the 0--1$\%$ (upper row) and $v_{2}^{\mathrm{AA}}$ extracted from \cite{Abelev:2014pua} (left) $v_{3}^{\mathrm{AA-MC}}$ (middle) and $v_{4}^{\mathrm{AA-MC}}$ (right) for the same particle species at the 20--30$\%$ (lower row) centrality ranges. This figure illustrates how $v_{2}(p_{\mathrm{T}})$, $v_{3}(p_{\mathrm{T}})$ and $v_{4}(p_{\mathrm{T}})$ develop for different particle species in the same centrality range. A clear mass ordering is seen in the low $p_{\mathrm{T}}$ region (for $p_{\mathrm{T}}<3$ GeV/$c$) for $v_{2}(p_{\mathrm{T}})$, $v_{3}(p_{\mathrm{T}})$ and $v_{4}(p_{\mathrm{T}})$, which rises from the interplay between the anisotropic flow harmonics and radial flow \cite{Huovinen:2001cy,Teaney:2000cw,Voloshin:1996nv,Shen:2011eg}. Radial flow creates a depletion in the particle spectrum at low $p_{\mathrm{T}}$ values, which increases with increasing particle mass and transverse velocity. When this effect is embedded in an environment where azimuthal anisotropy develops, it leads to heavier particles having smaller $v_{n}$ value compared to lighter ones at a given value of $p_{\mathrm{T}}$ \cite{Huovinen:2001cy,Teaney:2000cw,Voloshin:1996nv,Shen:2011eg}. \\
Furthermore, the $v_{n}(p_{\mathrm{T}})$ values show a crossing between pions, kaons and protons, that, depending on the centrality and the order of the flow harmonic, takes place at different $p_{\mathrm{T}}$ values. It is seen that the crossing between e.g. $\pi^{\pm}$ and $\mathrm{p}+\overline{\mathrm{p}}$ happens at lower $p_{\mathrm{T}}$ for mid-peripheral than for the most-central collisions. For the 0--1$\%$ centrality range, the crossing point moves to higher  $p_{\mathrm{T}}$ values for $v_2$, since the common velocity field, which exhibits a significant centrality dependence \cite{Abelev:2013vea}, affects heavy particles more. The current study shows that this occurs not only in the case of elliptic flow but also for higher flow harmonics (i.e. triangular and quadrangular flow). For higher values of $p_{\mathrm{T}}$ ( $p_{\mathrm{T}}>3$ GeV/$c$), particles tend to group according to their type, i.e. mesons and baryons,  however this grouping holds at best approximately, similarly to what was observed in \cite{Abelev:2014pua}.\\
\vspace{-1cm}
\section{Summary}
\label{Summary}
The first measurement of $v_{2}^{\mathrm{AA-MC}}(p_{\mathrm{T}})$, $v_{3}^{\mathrm{AA-MC}}(p_{\mathrm{T}})$ and $v_{4}^{\mathrm{AA-MC}}(p_{\mathrm{T}})$ for $\pi^{\pm}$, $\mathrm{K}^{\pm}$ and $\mathrm{p}+\overline{\mathrm{p}}$ for the 0--1$\%$ and 20--30$\%$ centrality ranges of Pb--Pb collisions at $\sqrt{s_{\mathrm{NN}}} = 2.76$ TeV were reported in this contribution. The second, third and fourth Fourier coefficient were calculated with the Scalar Product method, using a pseudo-rapidity gap of $|\Delta\eta| > 0$ between the identified hadron under study and each of the reference flow particles and applying a subtraction for non-flow contributions based on HIJING.\\ 

The higher flow harmonics (i.e. $v_3$ and $v_4$) become gradually larger than $v_2$ at around 1 ($v_3$) and 2.5 GeV/$c$ ($v_4$) in $p_{\mathrm{T}}$ for pions for the 1$\%$ most central Pb--Pb collisions. For heavier particles, the crossing points shift to higher $p_{\mathrm{T}}$ values due to the interplay between radial and azimuthal flow. A distinct mass ordering was found for both centralities in the low transverse momentum region i.e. for $p_{\mathrm{T}} < 3$ GeV/$c$, which again is attributed to a result of the interplay between the azimuthal anisotropy and radial flow. Finally, for $p_{\mathrm{T}} > 3$ GeV/$c$ the $v_{n}(p_{\mathrm{T}})$ for all harmonics tend to group according to their particle type at an approximate level.\\

\vspace{-0.75cm}




\bibliographystyle{elsarticle-num}
\bibliography{<your-bib-database>}



\end{document}